# Quantum Input-Output Theory for Optical Cavities with Arbitrary Coupling Strength: Application to Two-Photon Wave-Packet Shaping

M. G. Raymer[1] and C. J. McKinstrie[2]

[1] Department of Physics, University of Oregon, Eugene, Oregon 97403, USA
[2] Bell Laboratories, Alcatel-Lucent, Holmdel, New Jersey 07733, USA

We develop quantum-optical input-output theory for resonators with arbitrary coupling strength, and for input fields whose spectrum can be wider than the cavity free-spectral range, while ensuring that the field-operator commutator relations in space-time variables are correct. The cavity-field commutator exhibits a series of space-time 'echoes,' representing causal connections of certain space-time points by light propagation. We apply the theory to two-photon wave-packet shaping by cavity reflection, which displays a remarkable illustration of dispersion cancellation. We also show that the theory is amenable to inclusion of intracavity absorbing and emitting atoms, allowing, for example, dissipative losses within the cavity to be incorporated in a quantum mechanically correct way.

## 1. Introduction

Input-output theory for optical cavities or resonators plays a crucial role in quantum and classical optics because of the enhancement of coupling between external light fields and the cavity modes, as well as between the cavity modes and any medium inside the cavity. By input-output (I-O) theory, one means a differential equation of motion for the field in space-time variables, coupled with appropriate cavity boundary conditions, that is amenable to inclusion of intracavity absorbing and emitting atoms. The theory of optical cavities was well developed in the decades following the invention of the laser. Nevertheless, a simple formulation treating input-output theory as a scattering problem that is valid in quantum as well as classical contexts was not developed until the 1980s, when the ideas of quantum-noise squeezing and cavity-QED were being developed. Collett and Gardiner [1], Gardiner and Savage [2], and Yurke [3] developed a quantum mechanical linear-systems approach that describes the evolution of the input field, cavity field, and output field. That approach crucially ensures correct quantum commutator relations between the positive- and negative-frequency components of the electromagnetic field, or equivalently the photon annihilation and creation operators. This ensures that the field evolution is quantum mechanically unitary, as it must be in the absence of dissipative losses. The I-O theory has been generalized to include spatially complex cavity structures and intracavity losses. [4, 5]

The principal limitation of the most commonly used input-output theories [1, 6] is that they are restricted to the limit of high cavity quality-factor Q, in which the cavity mirrors or other coupling junctions have very high reflectivity, and no dissipative losses. In the high-Q limit, and assuming that no standing waves are formed, the optical field is uniformly distributed along the beam axis; that is, there are no pulse-propagation effects inside the cavity. For this limit to hold, the field must have spectral width much smaller than the cavity's free-spectral range (FSR). The field necessarily evolves negligibly during one round trip in the cavity. This high-Q, "good-cavity" limit is commonly used in many experimental situations, and so the input-output theory developed in the 1980s has seen widespread use. On the other hand, there are situations in which a cavity with lower mirror reflectivity, and thus lower Q, are used. For example, in a cavity with non-negligible dissipative loss, it may be advantageous to increase the cavity output coupling by decreasing the mirror reflectivity, in order to increase the efficiency of extracting field energy from the cavity. A more general theory is needed to describe such situations.

In this paper we develop input-output theory for arbitrary coupling-junction strength, and for input fields whose spectrum can be wider than the cavity FSR, while ensuring that the field-operator commutator relations are correct. A key development is expressing the fields and field commutators in the space-time variables, and extending previous results [7, 4] for these. In particular, we find that the cavity-field commutator exhibits a series of space-time 'echoes,' representing causal connections of certain space-time points by light propagation.

Figure 1 shows examples of cavity types being considered. Each has two input channels and two output channels, with the two channels propagating in opposite directions. The input-output coupling is created by a junction, which may be a mirror or a waveguide coupler. In the common case that only a single input channel is occupied by light, the system operates in traveling-wave geometry where standing waves do not form. In that case and in the absence of backscattering and nonlinear coupling, the two counter-propagating fields evolve independently of each other and so we can ignore one of them. (However, one should be aware that if there is a structure in the cavity medium that is capable of acting as a nonlinear coupling or as a diffraction grating, such as a gain medium periodically modulated on sub-wavelength scales, then the forward and backward waves are coupled and the solution must be generalized to account for this.)

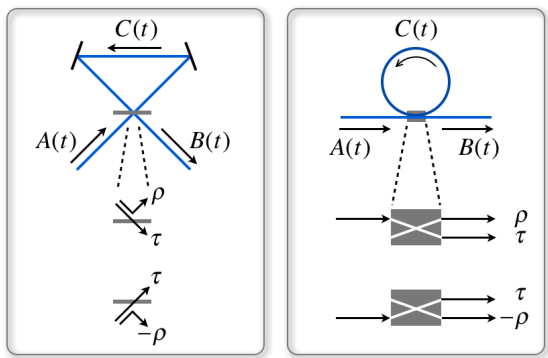

Fig. 1 (Color online) Ring cavity configurations with a single input, showing the conventions used for amplitude transmission coefficient $\tau$ and reflection coefficient $\rho$. We choose a phase convention where external reflection creates a phase flip, and internal reflection does not.

There is a close relationship between the present theory and laser theory, and we make use of this. In particular, early papers on the theory of multimode optical cavity instabilities and mode locking provide part of the inspiration for our formulation. [8, 9, 10, 11]

In this paper we present the general formalism for arbitrary coupling strength, expressed in both space-frequency and space-time domains. We apply it to the problem of two-photon wave-packet shaping by cavity reflection. We also show how dissipative losses within the cavity can be incorporated in a quantum mechanically correct way.

## 2. Input-Output and Commutation Relations

As mentioned above, it is simplest to consider only a single input beam and a single output beam, in which case the field in the cavity is a traveling-wave one with no standing-wave effects. It is straightforward to generalize to the case of two counter-propagating inputs and therefore standing waves in the cavity. In addition, only a single transverse mode is considered, although it is straightforward to include more such modes.

In the presence of a local electric-dipole polarization $P(z,t)$ (which equals zero for an empty cavity), the normalized electric field amplitude (or photon annihilation operator) $C(z,t)$ obeys the partial differential traveling-wave Maxwell equation:

$$(\partial_t + v\partial_z)C(z,t) = \alpha P(z,t) , \quad (1)$$

where $z$ is the distance traveled around the cavity path starting at the junction, and $v$ is the group velocity (speed of light $c$ for an empty cavity). $\alpha$ is a coupling parameter. The carrier wave has been factored from the field amplitude, and all frequencies are specified relative to the optical carrier frequency. The input coupling and periodicity of the cavity field are represented by the boundary condition:

$$C(0_+,t) = \rho\, C(L_-,t) + \tau A(t) \quad (2)$$

relating the cavity field to the input field $A(t)$ at the coupling junction. Here $0_+$ is the location inside the cavity immediately following the coupling junction and $L_-$ is the location inside the cavity just before impinging on the coupling junction. We choose a phase convention in which the coupling coefficients $\tau$ (for transmission) and $\rho$ (for reflection) are real and thus they must obey $\tau^2 + \rho^2 = 1$. This relation is the same as used in laser theory, [8, 9, 10, 11] but generalized to include the input field, which is normally not considered unless injection locking is being considered. [12]

The other boundary condition that must be satisfied determines the output field $B(t)$ in terms of the input field $A(t)$ and the cavity field:

$$B(t) = \tau C(L_-,t) - \rho A(t) . \quad (3)$$

It is notable that Eqs. (2) and (3) are the same as the standard beam-splitter relations [13], familiar in quantum optics. They can be written in common matrix form as

$$\begin{pmatrix} B(t) \\ C(0_+,t) \end{pmatrix} = \begin{pmatrix} \tau & -\rho \\ \rho & \tau \end{pmatrix} \begin{pmatrix} C(L_+,t) \\ A(t) \end{pmatrix} . \quad (4)$$

These relations enter the theory here as cavity-field boundary conditions rather than as the (sometimes) ad hoc relations that are postulated in order to maintain commutation relations in free space. [14] As in those treatments, the minus sign on $-\rho$ makes the matrix in Eq. (4) unitary. In fact, we will show that, rather than maintain free-space commutation relations, Eqs. (1) – (3) lead to significant alterations to the commutation relations in a way that maintains causality and unitarity in the theory.

The input field, being a freely propagating field in free space, obeys the familiar commutation relation (for a field whose bandwidth is significantly smaller than the optical carrier frequency):

$$[A(t), A^\dagger(t')] = \delta(t-t') . \quad (5)$$

This is a special case of the more general relation that applies to a field $A(z,t)$ freely propagating at speed $v$ in the absence of dispersion:

$$[A(t,z), A^\dagger(t',z')] = \delta(t-t'-(z-z')/v)) , \quad (6)$$

which results simply from $A(z,t) = A(0, t-z/v)$. This shows that the input field (assuming no dispersion) commutes with itself at all times, except those that are causally connected by the speed of light in free space. Physically, only when a commutator is zero are the two operators independently measurable.

The goal is to deduce from Eqs. (1) – (5) the cavity-field operator and the output-field operator, as well as their commutation relations. We first consider the case of an empty cavity, so that $P(t) = 0$. We introduce the Fourier-transform fields according to:

$$f(z,\omega) = \int_{-\infty}^{\infty} dt\, e^{i\omega t} F(z,t) . \quad (7)$$

(Throughout the paper we denote frequency-domain functions by lower-case letters.) Then Eqs. (1) – (5) imply

$$-i\omega c(z,\omega) + v\partial_z c(z,\omega) = 0 \quad (8)$$
$$c(0_+, \omega) = \rho\, c(L_-, \omega) + \tau a(\omega) \quad (9)$$
$$b(\omega) = \tau c(L_-, \omega) - \rho a(\omega) . \quad (10)$$

Equation (8) easily yields for the cavity field:

$$c(L_-, \omega) = c(0_+, \omega) \exp(i\omega T) , \quad (11)$$

where $T = L/v$ is the cavity round-trip time. Equation (11) represents a time shift; by transforming back to the time domain and using the Fourier-shift theorem:

$$C(L_-, t) = C(0_+, t-T) . \quad (12)$$

Inserting (11) into (9), and solving, yields

$$c(0_+, \omega) = \frac{\tau}{1-\rho \exp(i\omega T)} a(\omega) , \quad (13)$$

a result familiar from classical cavity theory. Equation (13) is written in a linear-response form by defining a Green function $G_{ca}(\omega)$

$$G_{ca}(\omega) = \frac{\tau}{1-\rho \exp(i\omega T)} , \quad (14)$$

so that

$$c(0_+, \omega) = G_{ca}(\omega) a(\omega) . \quad (15)$$

The modulus-square of the Green function $G_{ca}(\omega)$ is plotted in Fig. 2, as a reminder of the well-known enhancement of the density of states that occurs near the cavity resonances. The free-spectral range (FSR) of the cavity in radians per second is $\Omega = 2\pi/T$. (Throughout the paper time is measured in units of $T$, position in units of $L$, and other variables are scaled accordingly.) The integral of $|G_{ca}(\omega)|^2$ over any integer multiple of the free-spectral range yields the value 1, showing that the number of states is conserved, while the density of states is redistributed. This is related to an approach picturesquely called the 'modes of the universe.' [15] In terms of that picture, our cavity is embedded in a nearly infinite-length cavity, and the modes of that large cavity (which have very small mode frequency spacing) are 'pulled' toward the resonances of our cavity by the dispersion it induces (or equivalently by the boundary conditions). This leads to a 'piling-up' of mode density near the resonances.

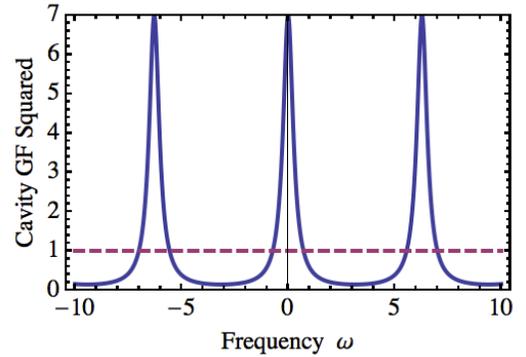

Fig.2 (Color online) Modulus-square of cavity Green function Eq. (14) versus frequency, for junction reflectivities $\rho = 0$ (dashed) and $\rho = 0.75$ (solid), and $T = 1$.

The output-field spectral amplitude $b(\omega)$ is given in terms of a different Green function $G_{ba}(\omega)$ as:

$$b(\omega) = G_{ba}(\omega) a(\omega) , \quad (16)$$

where

$$G_{ba}(\omega) = \exp(i\omega T)\left(\frac{1-\rho\exp(-i\omega T)}{1-\rho\exp(i\omega T)}\right) , \quad (17)$$
$$= \exp(i\theta(\omega))$$

from which we can see that in the frequency domain the output field differs from the input field only by a unit-magnitude factor with a frequency-dependent phase $\theta(\omega)$. This fact is consistent with energy conservation.

The inverse relation is easy to write; from Eqs. (16) and (17):

$$a(\omega) = (G_{ba}(\omega))^{-1} b(\omega)$$
$$= \exp(-i\theta(\omega))b(\omega) \quad . \quad (18)$$
$$= G_{ba}(\omega)^* b(\omega)$$

The inverse Green function is simply the complex conjugate of the forward one: $G_{ab}(\omega) = G_{ba}(\omega)^*$. As can be seen from (17), the complex conjugate corresponds to simply replacing $T$ by $-T$ in $G_{ba}(\omega)$.

The above relations are common lore in cavity theory, but to the best of our knowledge they have not previously been exploited to derive simple input-output relations for quantum fields. To further this goal, we derive the quantum-mechanical ramifications of the above relations.

The commutator for the input field is, in the frequency domain:

$$\left[a(\omega), a^\dagger(\omega')\right] = 2\pi\delta(\omega - \omega') \quad . \quad (19)$$

The commutator for the cavity field in the frequency domain is, from Eqs. (15) and (19),

$$\left[c(z,\omega), c^\dagger(z,\omega')\right] = \left|G_{ca}(\omega)\right|^2 2\pi\delta(\omega - \omega')$$
$$= \left|\frac{\tau}{1 - \rho\exp(i\omega T)}\right|^2 2\pi\delta(\omega - \omega') . \quad (20)$$

This commutator is different from the free-space one, and reflects the increase of the density of states near the cavity resonances.

The cavity field can be expressed in a different way using a Taylor-series expansion:

$$c(0_+, \omega) = \frac{\tau a(\omega)}{1 - \rho\exp(i\omega T)} = \tau a(\omega)\sum_{n=0}^{\infty}\rho^n e^{in\omega T} \quad . \quad (21)$$

Transforming back to the time domain yields

$$C(0_+, t) = \int_{-\infty}^{\infty}\frac{d\omega}{2\pi}e^{-i\omega t}\tau a(\omega)\sum_{n=0}^{\infty}\rho^n e^{in\omega T}$$
$$= \tau\int_{-\infty}^{\infty}dt' A(t')\sum_{n=0}^{\infty}\rho^n \delta(t - nT) \quad (22)$$
$$= \tau\sum_{n=0}^{\infty}\rho^n A(t - nT) .$$

The first term in the sum is the beam-splitter transfer function, while the remaining terms are delayed and attenuated replicas ('echoes') of the input. This, too, can be expressed in linear-response form, by introducing the time-domain Green function $\widetilde{G}_{ca}(t)$, which is the Fourier-transform of $G_{ca}(\omega)$. Then

$$C(0_+, t) = \int_{-\infty}^{\infty}dt' \widetilde{G}_{ca}(t - t')A(t') \quad , \quad (23)$$

where

$$\widetilde{G}_{ca}(t) = \tau\sum_{n=0}^{\infty}\rho^n \delta(t - nT) \quad . \quad (24)$$

The commutation relation for the cavity field in the time domain is derived in Appendix 1 from Eq. (22), and is found to be:

$$\left[C(0_+, t), C^\dagger(0_+, t')\right] =$$
$$= \tau^2\sum_{n=0}^{\infty}\sum_{m=0}^{\infty}\rho^m\rho^n\left[A(t - nT), A^\dagger(t' - mT)\right] \quad (25)$$
$$= \sum_{k=-\infty}^{\infty}\rho^{|k|}\delta(t - t' - kT) .$$

This shows that the cavity field at position $z = 0_+$ commutes with itself at all times except those separated by integer multiples of the cavity round-trip time; that is, at those times that are causally connected by the speed of light in the cavity. In the limit that there is no cavity, i.e. $\rho \to 0$, this recovers the free-space relation Eq. (5), as expected. For nonzero $\rho$ the commutator decays as $\rho^{|k|}$, indicating loss of memory or correlation between widely separated times.

The commutator in Eq. (25) can easily be generalized to account for different positions in the cavity in similar manner to Eq. (6) for the input field. Note that Eq. (1) with $P = 0$ implies $C(z,t) = C(0_+, t - z/v)$; then Eq. (25) implies

$$\left[C(z,t), C^\dagger(z',t')\right] = \sum_{k=-\infty}^{\infty}\rho^{|k|}\delta(t - z/v - (t' - z'/v) + kT) .$$
$$(26)$$

For equal times this becomes

$$[C(z,t),C^\dagger(z',t)] = \sum_{k=-\infty}^{\infty} \rho^{|k|}\delta((z'-z+kL)/v)$$
$$= \delta((z'-z)/v) \quad (27)$$

where the sum does not contribute because $z$ is contained within the interval $z \in [0,L]$. This agrees with the fundamental field commutator result verified in [7], where it was noted that Eq. (27) is the same inside and outside the cavity, as it must be. What that report left unsaid is that when different times are considered, as in Eq. (26), the commutator inside the cavity is *not* the same as that of the input field. The temporal evolution – here manifested as echoes – affects the commutator, consistent with causality.

Figure 3 illustrates the space-time structure of the commutator, from Eq. (26). Slanted white lines indicate values of $t$ and $z$ where the commutator is non-zero. For $t=0$ (horizontal white lines), it can be seen that as $z'$ increases from 0 to L, the spatial position where the commutator is nonzero (indicated by the line crossings) moves with $z'$. There is a single crossing for any fixed value of $z'$, in agreement with Eq. (27). In contrast, for fixed $z$, $z'$, and $t'$ values, there are an infinite number of times $t$ at which the commutator is non-zero.

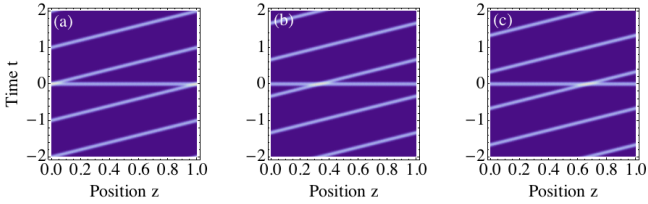

Fig. 3   (Color online) Modulus of the cavity-field commutator, versus $z$ and $t$, for $t' = 0$, and (a) $z' = 0$, (b) $z' = 0.333$, and (c) $z' = 0.666$. The cavity length $L=1$; the speed of light $v = 1$; time $t$ is measured in units $L/v$. Horizontal white lines indicate $t=0$ regions. Mirror reflectivity $\rho^2 = 0.998$.

We can also calculate the commutator between the cavity field at position $z = 0_+$ and the input field:

$$[C(0_+,t),A^\dagger(t')] = \tau \sum_{n=0}^{\infty} \rho^n [A(t-nT),A^\dagger(t')]$$
$$= \tau \sum_{n=0}^{\infty} \rho^n \delta(t-t'-nT). \quad (28)$$

The cavity field $C(0_+,t)$ commutes with the input field $A^\dagger(t')$ for all times $t < t'$ because later values of the input field cannot affect the cavity field at earlier times.

Causality is satisfied. This can also be generalized to account for different positions in the cavity, as in Eq. (26). The commutator between the cavity field and the output field can be found by similar means.

The output field $B(t)$, being a freely propagating field in free space, must obey the same commutation relations as does the input field, i.e., Eqs. (5) and (19). This is easy to see in the frequency domain, where the commutator, from Eqs. (16)–(19), is

$$[b(\omega),b^\dagger(\omega')] = |G_{ba}(\omega)|^2 2\pi\delta(\omega-\omega')$$
$$= 2\pi\delta(\omega-\omega'). \quad (29)$$

On the other hand, the mapping between the input and output fields is nontrivial when expressed in the time domain. To derive this, express the output-field Green function in the frequency domain, Eq. (17), as

$$G_{ba}(\omega) = -\rho + \left(\frac{\tau^2 \exp(i\omega T)}{1-\rho \exp(i\omega T)}\right)$$
$$= -\rho + \tau^2 \sum_{n=1}^{\infty} \rho^{n-1} e^{in\omega T}. \quad (30)$$

An inverse Fourier transform then gives the output Green function in the time domain:

$$\widetilde{G}_{ba}(t) = -\rho\delta(t) + \tau^2 \sum_{n=1}^{\infty} \rho^{n-1}\delta(t-nT), \quad (31)$$

Therefore, the output field in the time domain is:

$$B(t) = \int_{-\infty}^{\infty} dt' \widetilde{G}_{ba}(t-t')A(t'), \quad (32)$$

or

$$B(t) = -\rho A(t) + \tau^2 \sum_{n=1}^{\infty} \rho^{n-1} A(t-nT). \quad (33)$$

The first term here is the beam-splitter transfer function, while the terms in the sum are 'echoes.' It can be shown that Eq. (33) is consistent with Eqs. (3), (12), and (22). From this result, the commutation relation for the output field in the time domain can be derived as a consistency check, and indeed is found to be

$$[B(t),B^\dagger(t')] = \delta(t-t'), \quad (34)$$

that is, the same free-space commutator that is obeyed by the input field. The derivation is given in Appendix 2. The generalized form in Eq. (6) also holds for the output field. This reflects the fact that the output field, because it is traveling in free space, can in principle be measured with arbitrarily high precision simultaneously at distinct space-time points not connected by causal propagation.

The inverse relation can be written in the time domain as well; from Eq. (18), we saw that it corresponds to simply replacing $T$ by $-T$. Thus

$$A(t) = \int_{-\infty}^{\infty} dt' \widetilde{G}_{ab}(t-t') B(t') \;, \quad (35)$$

where

$$\widetilde{G}_{ab}(t) = -\rho \delta(t) + \tau^2 \sum_{n=1}^{\infty} \rho^{n-1} \delta(t+nT) \;, \quad (36)$$

so,

$$A(t) = -\rho B(t) + \tau^2 \sum_{n=1}^{\infty} \rho^{n-1} B(t+nT) \;. \quad (37)$$

This result can be verified explicitly by substituting Eq. (33) into Eq. (37) and performing the sums.

## 3. Reduction to Standard High-Q Input-Output Theory

To verify that the theory above includes the standard I-O theory [1] as a limiting case, we consider the limit regime satisfying three conditions:

*i*. The junction transmission coefficient $\tau$ is very small, so the cavity storage time is long.

*ii*. The cavity round-trip time $T$ is very small, compared to the duration of the input field pulse.

*iii*. The input field is narrow band, so it is well contained within a single FSR of the cavity.

For a transform-limited input field, conditions *ii* and *iii* are equivalent.

Define a cavity damping rate $\kappa$ by $\kappa = (1/T)\ln(1/\rho)$, so that $\rho = e^{-\kappa T}$, which gives two ways to write the attenuation factor suffered by the field on each trip around the cavity. Then we can write, without approximation,

$$\widetilde{G}_{ca}(t) = \sum_{n=0}^{\infty} \tau e^{-\kappa nT} \delta(t-nT) \;. \quad (38)$$

In order to consider how $\widetilde{G}_{ca}(t)$ behaves in the high-Q limit, note that it is a distribution (not a function), so it has meaning only as a factor inside an integral. The relevant integral is Eq. (23), which gives

$$C(0_+,t) = \sum_{n=0}^{\infty} \tau e^{-\kappa nT} A(t-nT) \;. \quad (39)$$

If the conditions *i - iii* are met, meaning that $\kappa T \ll 1$, then the sum in Eq. (39) can be well approximated by an integral:

$$C(0_+,t) \approx \int_{-\infty}^{t} dt' \frac{\tau}{T} e^{-\kappa(t-t')} A(t') \;, \quad (40)$$

where we used $t' = t - nT$ and noted that $n \geq 0$ implies $t' \leq t$. Comparing this to Eq. (23) shows that in this limit the Green function can be effectively replaced by

$$\widetilde{G}_{ca}(t-t')_{eff} = \frac{\tau}{T} e^{-\kappa(t-t')} \Theta(t-t') \;, \quad (41)$$

where $\Theta(x)$ is the Heaviside step (theta) function. Transforming this to the frequency domain gives

$$G_{ca}(\omega)_{eff} = \int_{-\infty}^{\infty} dt e^{i\omega t} \frac{\tau}{T} e^{-\kappa t} \Theta(t) = \frac{\tau/T}{\kappa - i\omega} \;, \quad (42)$$

a complex Lorentzian, as expected for a single, narrow resonance of a cavity. The same result is obtained directly by considering Eq. (14) in the limit $\omega T \to 0$, and using $\rho = e^{-\kappa T} \approx 1 - \kappa T \approx 1 - \tau^2/2$, which implies $\kappa \approx (1-\rho)/T$ and $\kappa \approx \tau^2/2T$, which is standard in high-Q cavity theory.

Near the resonance, the shape of the spectral response in Eq. (42) is similar to the exact result given by Eq. (14), so it might seem tempting to apply the approximate form even in the intermediate-loss regime, where $\rho$ is significantly different than 1. The problem with this idea is that the approximate form Eq. (42) has a maximum value $\tau/\ln(1/\rho)$, whereas the exact result has maximum value $\tau/(1-\rho)$. These agree only in the limit $\rho \to 1$, which is the high-Q regime. Therefore, we restrict application of Eq. (42) to the high-Q regime.

Standard I-O theory in the high-Q regime can be recovered easily by noting that the solution in Eq. (40) (valid in the limit $\omega T \to 0$) satisfies the following differential equation:

$$\partial_t C(0_+,t) = -\kappa C(0_+,t) + \frac{\tau}{T} A(t) \;. \quad (43)$$

This fundamental equation of motion for the cavity field is supplemented with the output-field equation, Eq. (3):

$$B(t) = \tau C(L_-,t) - \rho A(t) \;. \quad (44)$$

The goal of standard I-O theory is to be able to treat the cavity field as an effective single mode, called a 'quasimode,' with annihilation operator $C(t)$ that obeys the commutator $[C(t),C^\dagger(t)]=1$. To this end, we note that in the limit $\omega T \to 0$, the effect of one round trip is negligible, so Eq. (11) and (12) imply that $c(L_-,\omega) \approx c(0_+,\omega)$ and $C(L_-,t) \approx C(0_+,t)$, so we define $C(t) = \sqrt{T}\,C(0_+,t)$, where we also introduced a scaling factor $\sqrt{T}$. This makes the cavity field dimensionless. Then, also using $\sqrt{\kappa} \approx \sqrt{\tau^2/2T}$, we find

$$\partial_t C(t) = -\kappa C(t) + \sqrt{2\kappa}\,A(t) \ . \tag{45}$$

The output-field equation becomes, in the limit $\rho \to 1$

$$B(t) = \sqrt{2\kappa}\,C(t) - A(t) \ . \tag{46}$$

The commutator of the (rescaled) quasimode operator is easily found from Eq. (40) to be:

$$[C(t),C^\dagger(t')] = \exp(-\kappa|t-t'|) \ . \tag{47}$$

This reduces to $[C(t),C^\dagger(t)]=1$ for equal times, justifying the scaling factor that we used. Equations (45) - (47) are the standard I-O theory for high-Q cavities, originally derived using a master-equation method. [1] Note that the commutator Eq. (34) is exactly upheld even with the approximations made in arriving at Eqs. (45) and (47).

It is helpful to compare graphically the forms of the commutator in the exact and approximate theories, as in Fig. 4. For high junction reflectivity, $\rho = 0.97$, the approximate commutator (using $\kappa \approx (1-\rho)/T$) acts like an accurate envelope for the exact commutator, which is a sum of delta functions. However, for $\rho = 0.70$, the approximate result deviates significantly from the true envelope of the delta functions. By using the exact expression for the damping rate, $\kappa = (1/T)\ln(1/\rho)$, the approximate commutator can be made to decay at the exact same rate as the exact commutator. But then, as stated above, the magnitudes of the Green functions in the frequency domain do not agree quantitatively unless $(1-\rho) \ll 1$, which is the high-Q limit.

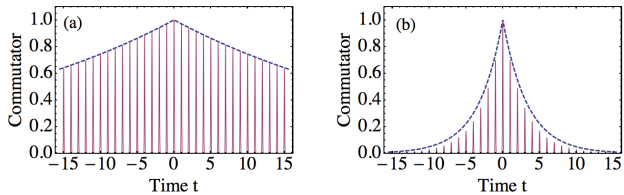

Fig. 4 (Color online) Cavity-field commutator $[C(t),C^\dagger(t')]$ versus time difference; solid - exact from Eq. (28); dashed - standard approximation from Eq. (47), using $\kappa \approx (1-\rho)/T$. (a) $\rho = 0.97$, (b) $\rho = 0.70$. (Delta functions are represented by narrow Gaussians for visualization.)

## 4. Cavity Shaping of Time-Frequency Entangled Two-Photon Wave-Packets

A standard example of non-classical light is the time-frequency-entangled photon pair. [16, 17, 18, 19] It can exhibit violations of Bell inequalities [16], violation of classical Maxwell electromagnetic theory [20, 21], and is useful in quantum cryptographic key distribution [22], among other applications. It is easily created using spontaneous parametric down-conversion in crystals [23, 24] or spontaneous four-wave mixing in fibers [25, 26, 27, 28], and since it contains only two photons is fully characterized by its fourth-order electric-field correlation function,

$$f(t_1,t_2) = \langle\Psi|A^\dagger(t_1)A^\dagger(t_2)A(t_2)A(t_1)|\Psi\rangle \ , \tag{48}$$

where the field operator is (if the light is not too broad band)

$$A(t) = \int \frac{d\omega}{2\pi} a(\omega) e^{-i\omega t} \ . \tag{49}$$

The state can be expressed equivalently in the frequency or time domains as

$$|\Psi\rangle = (2\pi)^{-2}\int d\omega \int d\omega'\, \widetilde{\psi}(\omega,\omega') a^\dagger(\omega)a^\dagger(\omega')|vac\rangle$$
$$= \int dt \int dt'\, \psi(t,t') A^\dagger(t)A^\dagger(t')|vac\rangle, \tag{50}$$

where $\psi(t,t')$ is the double Fourier transform of $\widetilde{\psi}(\omega,\omega')$ and the field operators obey the commutators Eq. (19) or Eq. (5). (Note that we are working in the Heisenberg picture, where the state $|\Psi\rangle$ is time independent.) The modulus-squared of the two-photon probability amplitude $|\psi(t_1,t_2)|^2$ gives the joint probability to detect photons at both times $t_1$ and $t_2$, and is determined by the properties of the down-conversion crystal and the laser field used to pump it. [29, 30, 31] Likewise, $|\widetilde{\psi}(\omega,\omega')|^2$ gives the joint spectral density – the probability to detect photons at both frequencies $\omega$ and $\omega'$.

The correlation function in Eq. (48) is the inner product of $A(t_2)A(t_1)|\Psi\rangle$ with its hermitian conjugate. So we evaluate:

$$A(t_2)A(t_1)|\Psi\rangle = \int dt \int dt' \psi(t,t') A(t_2) A(t_1) A^\dagger(t) A^\dagger(t') |vac\rangle$$
$$= (\psi(t_1,t_2) + \psi(t_2,t_1))|vac\rangle ,$$
(51)

where we used the commutator Eq. (5) repeatedly inside the integral to put the operators into normal order (annihilation operators to the right). Then the correlation function becomes $f(t_1,t_2) = |\Phi(t_1,t_2)|^2$, where we defined the 'two-photon wave function' as

$$\Phi(t_1,t_2) = \langle vac|A(t_2)A(t_1)|\Psi\rangle$$
$$= \psi(t_1,t_2) + \psi(t_2,t_1) .$$
(52)

This definition is standard notation in the quantum-field theory of massive particles. In quantum optics, it can be thought of as simply a function that is proportional to the two-photon detection amplitude [32, 33] or as a true photon wave function. [34, 35] Note that this function automatically has the correct boson symmetry under photon label exchange $(t_1 \leftrightarrow t_2)$. [35] This function can be engineered to be very narrow in the time difference $t_1 - t_2$, which implies that in the frequency domain there is strong anticorrelation between observed frequencies, with their sum equaling that of the pump laser. [36]

As an example of the input-output theory, consider what happens when such a two-photon state is incident on a cavity of the type in Fig.1. One might expect one of a few possibilities: the tight temporal correlation will be disrupted because a given photon in the pair may take any number of round trips around the cavity before emerging; the tight temporal correlation will be maintained because the effect of the cavity is only to introduce dispersion, and it is known that dispersion in bulk-media propagation is cancelled in certain situations for time-frequency entangled photon pairs [37]; or some combination of these two effects might occur.

The quantum state, Eq. (50), when expressed in the output-mode variables in the frequency domain, can be found using the inverse relation Eq. (18) to write $a^\dagger(\omega) = G_{ba}(\omega) b^\dagger(\omega)$. Then

$$|\Psi\rangle = (2\pi)^{-2} \int d\omega \int d\omega' \{\widetilde{\psi}(\omega,\omega') G_{ba}(\omega) G_{ba}(\omega')\}$$
$$\times b^\dagger(\omega) b^\dagger(\omega') |vac\rangle .$$
(53)

The quantity in brackets inside the integral is the output wave function. Because the green function $G_{ba}(\omega)$ is unimodular, this confirms that there is no change of the joint spectral density as a consequence of passing through the cavity.

The output state in the time domain is found by writing Eq. (35) for the creation operator (note the Green function is real):

$$A^\dagger(t) = \int_{-\infty}^{\infty} dt' \widetilde{G}_{ab}(t-t') B^\dagger(t') ,$$
(54)

and $\widetilde{G}_{ab}(t-t')$ is given by Eq.(36). Then Eq. (50) implies

$$|\Psi\rangle = \int dt \int dt' \psi(t,t') \int_{-\infty}^{\infty} d\tau \widetilde{G}_{ab}(t-\tau) B^\dagger(\tau)$$
$$\times \int_{-\infty}^{\infty} d\tau' \widetilde{G}_{ab}(t'-\tau') B^\dagger(\tau') |vac\rangle$$
$$= \int_{-\infty}^{\infty} d\tau \int_{-\infty}^{\infty} d\tau' \psi_{out}(\tau,\tau') B^\dagger(\tau) B^\dagger(\tau') |vac\rangle ,$$
(55)

where the two-photon amplitude at the output is

$$\psi_{out}(\tau,\tau') = \int dt \int dt' \widetilde{G}_{ab}(t-\tau) \widetilde{G}_{ab}(t'-\tau') \psi(t,t') .$$
(56)

The correlation function at the output is

$$f_{out}(t_1,t_2) = \langle \Psi | B^\dagger(t_1) B^\dagger(t_2) B(t_2) B(t_1) | \Psi \rangle ,$$
(57)

which equals the inner product of $B(t_2)B(t_1)|\Psi\rangle$ with its hermitian conjugate. In Appendix 3 we show that

$$B(t_2)B(t_1)|\Psi\rangle =$$
$$\int dt \int dt' \widetilde{G}_{ab}(t-t_1) \widetilde{G}_{ab}(t'-t_2) \Phi(t,t') |vac\rangle .$$
(58)

The correlation function for the output is thus $f_{out}(t_1,t_2) = |\Phi_{out}(t_1,t_2)|^2$, where the two-photon wave function is, using Eq. (36),

$$\Phi_{out}(t_1,t_2) = \int dt \int dt' \widetilde{G}_{ab}(t-t_1) \widetilde{G}_{ab}(t'-t_2) \Phi(t,t')$$
$$= \rho^2 \Phi(t_1,t_2) - \tau^2 \sum_{m=1}^{\infty} \rho^m \Phi(t_1, t_2 - mT)$$
$$- \tau^2 \sum_{n=1}^{\infty} \rho^n \Phi(t_1 - nT, t_2)$$
$$+ \tau^4 \sum_{n=1}^{\infty} \sum_{m=1}^{\infty} \rho^{n+m-2} \Phi(t_1 - nT, t_2 - mT) .$$
(59)

This clearly holds the possibility for photon-counting coincidences to occur at any combinations of delays suffered separately by the two photons. But two-photon quantum interference can eliminate some of these possibilities under certain conditions.

As an example, first consider the common case of a stationary down-conversion source, pumped by a constant (cw) laser field. In this case the two-time wave function depends only on the time difference, $\Phi(t_1,t_2) = D(t_1-t_2)$, with the width of the function $D$ being the coherence time. This implies a two-time wave function in the frequency domain proportional to

$$\varphi(\omega,\omega') \propto \delta(\omega+\omega')\int d\tau\, e^{i(\omega-\omega')\tau} D(\tau) , \quad (60)$$

showing the perfect frequency anticorrelation that is characteristic of this form of time-frequency entanglement. To evaluate the two-time wave function of the output field in this case, we use the math relation

$$\sum_{n=1}^{\infty}\sum_{m=1}^{\infty} \rho^{n+m} D(t_1-t_2+(n-m)T)$$
$$= \sum_{k=-\infty}^{\infty} \sum_{s=|k|/2+1}^{\infty} \rho^{2s} D(t_1-t_2+kT) \quad (61)$$
$$= \frac{\rho^2}{1-\rho^2}\sum_{k=-\infty}^{\infty} \rho^{|k|} D(t_1-t_2+kT) .$$

Then we find

$$\Phi_{out}(t_1,t_2) = \rho^2 D(t_1-t_2) - \tau^2 \sum_{m=1}^{\infty}\rho^m D(t_1-t_2+mT)$$
$$- \tau^2 \sum_{n=1}^{\infty}\rho^n D(t_1-t_2-nT)$$
$$+ \frac{\tau^4}{\rho^2}\sum_{n=1}^{\infty}\sum_{m=1}^{\infty} \rho^{n+m} D(t_1-t_2-(n-m)T)$$
$$= D(t_1-t_2). \quad (62)$$

The output field has the same narrow temporal correlation function as does the input field! The photons of a pair emerge together after scattering from the cavity.

To see the origin of this result, note that the second term canceled with the positive-$k$ parts of the fourth term (written in the form of Eq. (61)), whereas the third term canceled with the negative-$k$ parts of the fourth term. The $k=0$ part of the fourth term combined with the first term to yield the result. That is, quantum amplitudes with zero cavity transits for one photon and $m$ transits for the other (second and third terms) are cancelled by amplitudes with $k$ transits for one photon and $k+m$ for the other. This is a remarkable example of the cancellation of dispersion that is known for two-photon light with perfect frequency anticorrelation. [37]

The case of a non-stationary two-photon source is also of interest. This occurs if the pump field is pulsed. In a special case we can model the wave function as a two-dimensional Gaussian, with parameter $\sigma$ giving the correlation time, and a second parameter $\beta$ giving the pulse duration,

$$\Phi(t_1,t_2) = \exp[-(t_1+t_2)^2/2\beta^2]\exp[-(t_1-t_2)^2/2\sigma^2] . \quad (63)$$

Then, from Eq. (59) and Eq. (63),

$$\Phi_b(t_1,t_2) = \left(\tau^2 F_0 + \rho^2 \exp[-(t_1+t_2)^2/2\beta^2]\right)$$
$$\times \exp[-(t_1-t_2)^2/2\sigma^2]$$
$$+ \tau^2 \sum_{m=1}^{\infty}\left(F_m - \rho^m \exp[-(t_1+t_2-mT)^2/2\beta^2]\right)$$
$$\times \exp[-(t_1-t_2+mT)^2/2\sigma^2] \quad (64)$$
$$+ \tau^2 \sum_{m=1}^{\infty}\left(F_m - \rho^m \exp[-(t_1+t_2-mT)^2/2\beta^2]\right)$$
$$\times \exp[-(t_1-t_2-mT)^2/2\sigma^2],$$

where

$$F_m = \tau^2 \sum_{s=|m|/2+1}^{\infty} \rho^{2s-2} \exp[-(t_1+t_2-2sT)^2/2\beta^2] . \quad (65)$$

The function $F_k$ goes to $\rho^{|k|}$ in the limit $\beta \to \infty$, thus recovering the result Eq.(62).

Consider the case of equal $\sigma$ and $\beta$, so the wave function of the input field is separable (expressible as a product of two function, one in $t_1$ and one in $t_2$). The magnitude of the two-photon wave function, Eq. (64) is plotted in Fig. 5.

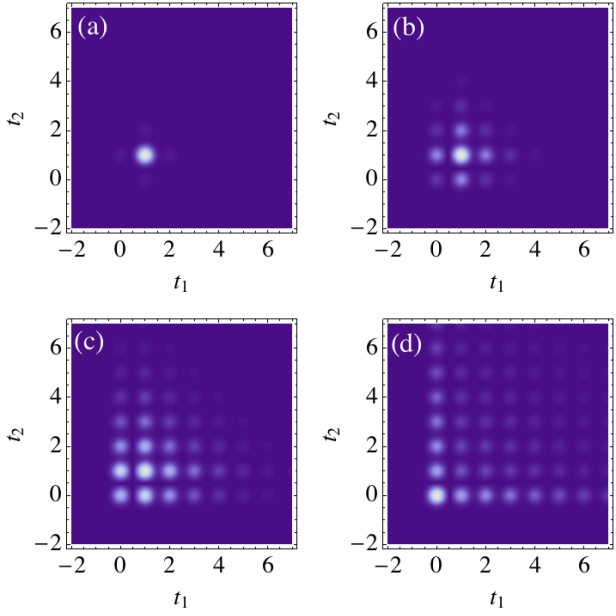

Fig. 5 (Color online) Magnitude of two-photon wave function $|\Phi_b(t_1,t_2)|$ vs. $t_1$ and $t_2$, from Eq.(64), for parameter values $\sigma = 0.3, \beta = 0.3$, $T = 1$, and the junction transmission varied as: (a) $\tau = 0.999$, (b) $\tau = 0.95$, (c) $\tau = 0.85$, (d) $\tau = 0.60$.

For $\tau = 0.999$ the dominant correlation peak occurs at $(t_1,t_2) = (1,1)$ because the light takes one full round trip in the nearly nonreflecting cavity before emerging. For $\tau = 0.60$ the dominant correlation peak occurs at $(t_1,t_2) = (0,0)$ because the light reflects from the junction, without delay, into the output beam. The output wave function is separable, as it was at the input. See Appendix 4. (This can be understood by noting that the output wave function in the frequency domain, in Eq. (53) retains its separability if the input state is separable. And by noting that separability in frequency implies separability in time.)

Figure 6 shows a case in which the input field's wave function is non-separable. The wave function retains this non-separability at the output, as it develops 'echoes.'

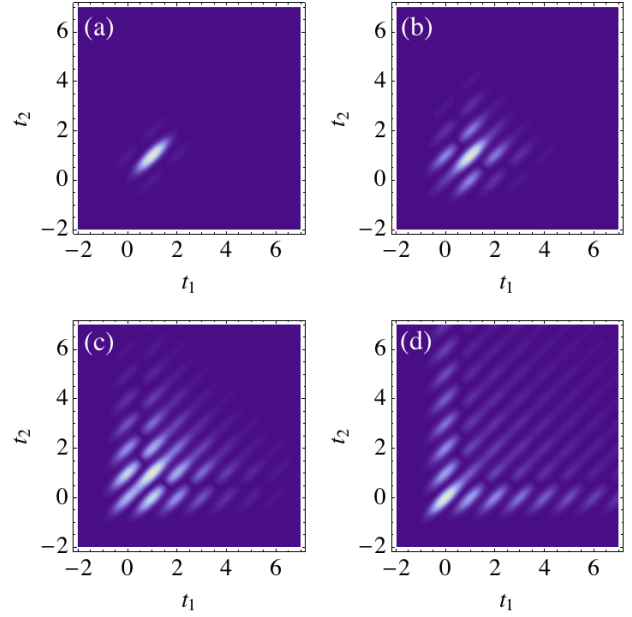

Fig. 6 (Color online) Magnitude of two-photon wave function $|\Phi_b(t_1,t_2)|$ vs. $t_1$ and $t_2$, from Eq.(64), for parameter values $\sigma = 0.2, \beta = 0.7$, $T = 1$, and the junction transmission varied as: (a) $\tau = 0.999$, (b) $\tau = 0.95$, (c) $\tau = 0.85$, (d) $\tau = 0.60$.

## 5. Inclusion of Dissipative Loss

To include dissipative loss in the arbitrary-coupling-strength model, we introduce an absorbing molecular medium throughout the cavity. Then Eq. (1) reads:

$$\begin{aligned}(\partial_t + v\partial_z)C(z,t) &= \alpha P(z,t) \\ \partial_t P(z,t) &= -\gamma P(z,t) - \beta C(z,t) + F(z,t),\end{aligned} \quad (66)$$

where $\alpha, \beta$ are coupling parameters, $\gamma$ is the damping rate for the molecular electric dipole polarization $P(z,t)$, and $F(z,t)$ is a quantum Langevin fluctuation operator obeying the commutator [38]

$$\left[F(z,t), F^\dagger(z',t')\right] = 2\gamma\,\delta(t-t')\delta(z-z'). \quad (67)$$

Integrate the equation for the dipole polarization, assuming the molecular damping is fast, making the absorber broad band:

$$\begin{aligned}P(z,t) &= \int_{-\infty}^t dt'\, e^{-\gamma(t-t')}\left(-\beta C(z,t') + F(z,t')\right) \\ &\approx -(\beta/\gamma)C(z,t) + F_P(z,t),\end{aligned} \quad (68)$$

where the effective Langevin fluctuation operator for the dipole polarization is

$$F_P(z,t) \equiv \int_{-\infty}^{t} dt' e^{-\gamma(t-t')} F(z,t') . \quad (69)$$

Calculate the commutator for the dipole fluctuation operator:

$$\begin{aligned}\left[F_P(z,t), F_P^\dagger(z',t')\right] &= \exp(-\gamma |t-t'|)\delta(z-z') \\ &\to 2\gamma \delta(t-t')\delta(z-z'),\end{aligned} \quad (70)$$

where the final step results from assuming the molecular damping is fast. That is, in the limit we consider, one can idealize the $P$-fluctuations as delta-correlated. Absorber models were introduced previously in I-O theory. [4, 5] Here it leads to the modified cavity-field propagation equation,

$$(\partial_t + v\partial_z)C(z,t) = -(\alpha\beta/\gamma)C(z,t) + \alpha F_P(z,t) , \quad (71)$$

where $\alpha\beta/\gamma$ plays the role of the attenuation rate of the cavity field. The key point is that dissipative loss always brings with it additional fluctuations, and these are accounted for by the Langevin operator.

It is straightforward to solve Eq. (71), along with the boundary conditions Eqs. (2) and (3), in steady state to study the effects of attenuation and fluctuation on the cavity and output fields. We leave this as an exercise.

**6. Discussion**

The main results of this study are: (1) Showing that the standard classical field propagation equations, Eqs. (1)-(3), provide a proper quantum mechanical description of input, cavity and output fields when the input coupling strength takes on arbitrary values; (2) Deriving explicit formulas for Green functions and commutators for the three fields in both space-frequency and space-time domains; (3) Confirming that the commutator Eq. (26) agrees with the fundamental equal-time field commutator Eq. (27), which must always be respected, and (4) Deriving the effects of a reflecting cavity on a two-photon wave-packet state.

The I-O theory formulated here is amenable to inclusion of intracavity absorbing and emitting atoms. Given that the equal-time field commutator agrees with the fundamentally required one Eq. (27) [7], the effects of atoms in the cavity may be accounted for by using the standard minimal-coupling atom-field interaction Hamiltonian. The theory can also account for dynamical absorbing media in the cavity by generalizing the equation of motion for the dipole polarization $P(z,t)$ in Eq. (66) to include multilevel media with or without population inversion and/or coherent control fields. This might be useful, for example, as a model for a quantum memory, and will be considered in a following paper.

We thank Ian Walmsley, Josh Nunn, Steven van Enk, Kartik Srinivasan, and an anonymous reviewer for helpful comments. MGR was supported by the National Science Foundation, EPMD (ENG) and AMOP (Physics).

**Appendix 1: Commutator for cavity field**

To verify Eq. (25), write

$$\begin{aligned}\left[C(0_+,t), C^\dagger(0_+,t')\right] &= \\ &= \tau^2 \sum_{n=0}^{\infty}\sum_{m=0}^{\infty} \rho^m \rho^n \left[A(t-nT), A^\dagger(t'-mT)\right] \\ &= \tau^2 \sum_{n=0}^{\infty}\sum_{m=0}^{\infty} \rho^{n+m}\delta(t-t'-(n-m)T) .\end{aligned} \quad (A1)$$

For $t \geq t'$ it is required that $n \geq m$; so

$$\begin{aligned}\left[C(0_+,t), C^\dagger(0_+,t')\right]_{t\geq t'} &= \\ &= \tau^2 \sum_{n=0}^{\infty}\sum_{m=0}^{n} \rho^{n+m}\delta(t-t'-(n-m)T) \\ &= \tau^2 \sum_{k=0}^{\infty}\sum_{s=0}^{\infty} \rho^{k+2s}\delta(t-t'-kT) \\ &= \sum_{k=0}^{\infty} \rho^k \delta(t-t'-kT),\end{aligned} \quad (A2)$$

where we used $\tau^2 + \rho^2 = 1$ and the general relation

$$\sum_{n=0}^{\infty}\sum_{m=0}^{n} f(n+m)g(n-m) = \sum_{k=0}^{\infty}\sum_{s=0}^{\infty} f(k+2s)g(k) . \quad (A3)$$

For $t \leq t'$, it is required that $n \leq m$; so

$$\left[C(0_+,t), C^\dagger(0_+,t')\right]_{t\leq t'} = \sum_{k=-\infty}^{0} \rho^{-k}\delta(t-t'+kT) . \quad (A4)$$

Combining the two cases gives:

$$\begin{aligned}\left[C(0_+,t), C^\dagger(0_+,t')\right] &= \\ &= \sum_{k=-\infty}^{\infty} \rho^{|k|}\delta(t-t'+kT) = \sum_{k=-\infty}^{\infty} \rho^{|k|}\delta(t-t'-kT) .\end{aligned} \quad (A5)$$

## Appendix 2: Commutator for output field

To verify Eq. (34), use Eqs. (2) and (3) to write

$$B(t) = -\frac{1}{\rho}A(t) + \frac{\tau}{\rho}C(0_+,t) \ . \quad (A6)$$

Use this and Eq. (28) to write

$$\begin{aligned}
&[B(t), B^\dagger(t')] = \\
&= \frac{1}{\rho^2}[A(t), A^\dagger(t')] - \frac{\tau}{\rho^2}[C(0_+,t), A^\dagger(t')] \\
&\quad - \frac{\tau}{\rho^2}[A(t), C^\dagger(0_+,t')] + \frac{\tau^2}{\rho^2}[C(0_+,t), C^\dagger(0_+,t')] \\
&= \frac{1}{\rho^2}\delta(t-t') - \frac{\tau^2}{\rho^2}\sum_{n=0}^{\infty}\rho^n\delta(t-t'-nT) \\
&\quad - \frac{\tau^2}{\rho^2}\sum_{n=0}^{\infty}\rho^n\delta(t-t'+nT) + \frac{\tau^2}{\rho^2}\left(\sum_{k=-\infty}^{\infty}\rho^{|k|}\delta(t-t'-kT)\right) \\
&= \frac{1}{\rho^2}\delta(t-t') - \frac{\tau^2}{\rho^2}\left(\delta(t-t') + \sum_{n\neq 0}\rho^n\delta(t-t'-nT)\right) \\
&\quad - \frac{\tau^2}{\rho^2}\left(\delta(t-t') + \sum_{n\neq 0}\rho^n\delta(t-t'+nT)\right) \\
&\quad + \frac{\tau^2}{\rho^2}\left(\delta(t-t') + \sum_{n\neq 0}\rho^n\delta(t-t'-nT) + \sum_{k\neq 0}\rho^{|k|}\delta(t-t'+kT)\right) \\
&= \left(\frac{1}{\rho^2} - \frac{\tau^2}{\rho^2}\right)\delta(t-t') = \delta(t-t')
\end{aligned}$$

(A7)

## Appendix part 3: Two-photon correlation function

To verify Eq. (58), write

$$\begin{aligned}
&B(t_2)B(t_1)|\Psi\rangle = \\
&= \int_{-\infty}^{\infty}d\tau\int_{-\infty}^{\infty}d\tau' F(\tau,\tau')B(t_2)B(t_1)B^\dagger(\tau)B^\dagger(\tau')|vac\rangle
\end{aligned} \quad (A8)$$

where

$$F(\tau,\tau') = \int dt\int dt'\widetilde{G}_{ab}(t-\tau)\widetilde{G}_{ab}(t'-\tau')\psi(t,t') \ . \quad (A9)$$

Repeated use of commutators gives

$$B(t_2)B(t_1)|\Psi\rangle = (F(t_1,t_2) + F(t_2,t_1))|vac\rangle \quad (A10)$$

The input two-photon function is:

$$\Phi(t_1,t_2) = \langle vac|A(t_2)A(t_1)|\Psi\rangle = \psi(t_1,t_2) + \psi(t_2,t_1) \quad (A11)$$

And the output one is:
$$\begin{aligned}
&\Phi_{out}(t_1,t_2) = \langle vac|B(t_2)B(t_1)|\Psi\rangle = \\
&= \int dt\int dt'\left(\widetilde{G}_{ab}(t-t_1)\widetilde{G}_{ab}(t'-t_2) + \widetilde{G}_{ab}(t-t_2)\widetilde{G}_{ab}(t'-t_1)\right)\psi(t,t')
\end{aligned}$$

(A12)

This can be written as

$$\Phi_{out}(t_1,t_2) = \int dt\int dt'\widetilde{G}_{ab}(t-t_1)\widetilde{G}_{ab}(t'-t_2)(\psi(t,t') + \psi(t',t))$$

(A13)

## Appendix 4: Separability of two-photon state

Proof that if the state is separable at the input, then it is separable at the output: If $\Phi(t,t') = \phi_1(t)\phi_2(t')$, then from Eq. (59),

$$\Phi_{out}(t_1,t_2) = \psi_{1out}(t_1)\psi_{2out}(t_2) \quad (A14)$$

where

$$\begin{aligned}
\psi_{1out}(t) &= \int dt'\widetilde{G}_{ab}(t'-t)\phi_1(t) \\
&= -\rho\phi_1(t) + \frac{\tau^2}{\rho}\sum_{n=1}^{\infty}\rho^n\phi_1(t-nT) \ , \\
\psi_{2out}(t) &= \int dt'\widetilde{G}_{ab}(t'-t)\phi_2(t') \\
&= -\rho\phi_2(t) + \frac{\tau^2}{\rho}\sum_{n=1}^{\infty}\rho^n\phi_2(t-nT)
\end{aligned} \quad (A15)$$

Plotting this form Eq. (A14) for the example shown in Fig. 5 gives results identical to those shown there.